# Wafer-scale uniformity improvement of Dolan-bridge Josephson junction by shadow evaporation bias correction

Daria A. Moskaleva[1,2,3], Nikita D. Korshakov[1,2,3], Dmitry O. Moskalev[1,2], Anastasiya A. Solovyova[1], Alexey R. Matanin[1,2], Elizaveta I. Malevannaya[1,2], Nikita S. Smirnov[1], Maksim I. Teleganov[1], Yuri V. Panfilov[1] and Ilya A. Rodionov[1,2]

[1] FMN Laboratory, Bauman Moscow State Technical University, Moscow, 105005, Russia
[2] Dukhov Automatics Research Institute, VNIIA, Moscow, 127030, Russia
[3] These authors contributed equally: Daria A. Moskaleva and Nikita D. Korshakov

*Electronic mail: irodionov@bmstu.ru

One of the practical limitations of solid-state superconducting quantum processors technology is frequency crowding due to low qubits fabrication reproducibility. Josephson junction 100 nm-scale nonlinear inductance of the qubits still suffers from Dolan-bridge shadow evaporation process. Here, we report on a robust wafer-scale Al/AlOx/Al Dolan-bridge Josephson junction (JJ) process using preliminary shadow evaporation bias resist mask correction and comprehensive oxidation optimization. We introduce topology correction model for two-layer resist mask biasing at a wafer-scale, which takes into account an evaporation source geometry. It results in Josephson junction area variation coefficient ($CV_A$) improvement down to 1.1% for the critical dimensions from 130×170 nm$^2$ to 130×670 nm$^2$ over 70×70 mm$^2$ (49 cm$^2$) wafer working area. Next, we investigate JJ oxidation process (oxidation method, pressure and time) and its impact on a room temperature resistance reproducibility. Finally, we combine both shadow evaporation bias correction and oxidation best practices for 4-inch wafers improving $CV_{R_N}$ down to 6.0/5.2/4.1% for 0.025 μm$^2$ JJ area and 4.0/3.4/2.3% for 0.090 μm$^2$ JJ area for 49/25/16 cm$^2$ wafer working area correspondingly. The proposed model and oxidation method can be useful for robust wafer-scale superconducting quantum processors fabrication.

## INTRODUCTION

There are several promising quantum computing platforms including trapped ions [1, 2], silicon photonics [3], semiconductor quantum dots [4, 5], NV center in diamonds [6, 7], superconducting quantum bits [8-11], etc. Quantum processors based on superconducting artificial atoms (Josephson junction qubits) are distinguished by good scalability and control. There are several superconducting NISQ processors have been demonstrated recently, including 54-qubits Sycamore quantum processor (Google) [12], 127-qubit Eagle and 1121-qubit Condor processors (IBM) [13, 14] and others [15-17]. However, qubits frequency crowding is still an issue for quantum processors scalability especially for fixed-frequency transmon qubits platform. One has to detune qubits resonance frequencies after quantum processors fabrication and cryogenic characterization. For example, in order to scale up beyond 1000 superconducting qubits with a cross-resonance gate architecture [18], less than 6 MHz qubit frequencies standard deviation is required. Qubit resonance frequency collisions (crowding) lead to bad energy levels design, increased quantum gates time and negative factors decreasing quantum processors gates fidelity [19, 20]. From the other hand, high-fidelity readout of superconducting quantum processors requires broadband quantum-limited cryogenic parametric amplifiers. Two well-known approaches in the field are Impedance-Matched Parametric Amplifiers (IMPA) [21-23] and Josephson Travelling Wave Parametric amplifier (JTWPA) [24, 25]. JTWPA are also very demanding to Josephson junction parameters reproducibility as they consist of hundreds identical elementary cells (LC-C oscillators) involved in a parametric amplification process. JTWPA ensure an exponentially increasing gain in case of high (98%) elementary cells electrical parameters reproducibility [26, 27].

Josephson junction area and tunnel barrier thickness variations are the key factors influencing superconducting qubits resonance frequency. There are two common methods for Josephson junction evaporation called Dolan bridge technique and Manhattan style process. Manhattan process doesn't require a fragile suspended resist bridge and less sensitive to resist thickness variation that allows increasing JJ area reproducibility. But Dolan bridge technique does not require large evaporation angles, unlike Manhattan junctions, which leads to improved line edge roughness. Moreover, it allows fabricating array of series-connected junctions for fluxonium qubit and quantum-limited parametric amplifier applications [9, 23]. For example, decreasing of room temperature resistance variation coefficient to $CV_{R_N}$= 2.3% [28, 29] is demonstrated for 39 mm$^2$ chip for 0.015…3.27 μm$^2$ Josephson junction areas. Recently, there were several techniques proposed to improve room temperature resistance and area variation coefficients over a wafer (Table 1). The best achieved reproducibility for Dolan bridge Josephson junctions is $CV_{R_N}$ down to 0.8 – 3.7% for 0.02 – 0.08 μm$^2$ areas inside 18 cm$^2$ working area on 4-inch wafers [30]. For Manhattan style junctions $CV_{R_N}$= 3.5% [31] and $CV_{R_N}$= 7.0% [32] is achieved for 49 cm$^2$ and 52 cm$^2$ working areas respectively. These results were achieved with both optimized Josephson junction fabrication and a novel multilayer fabrication route. However, the proposed processes cannot ensure both high JJ parameters reproducibility and high wafer filling-factor. With a post fabrication laser-annealing of each superconducting qubits [18, 33, 34] a variation coefficient $CV_{R_N}$ can be further improved down to 1% over a standard chip. However, it requires special equipment and it works for single Josephson junctions only (highly integrated junctions or JJ arrays are difficult to treat in a control way due to non-local nature of laser annealing with a few microns impact radius).

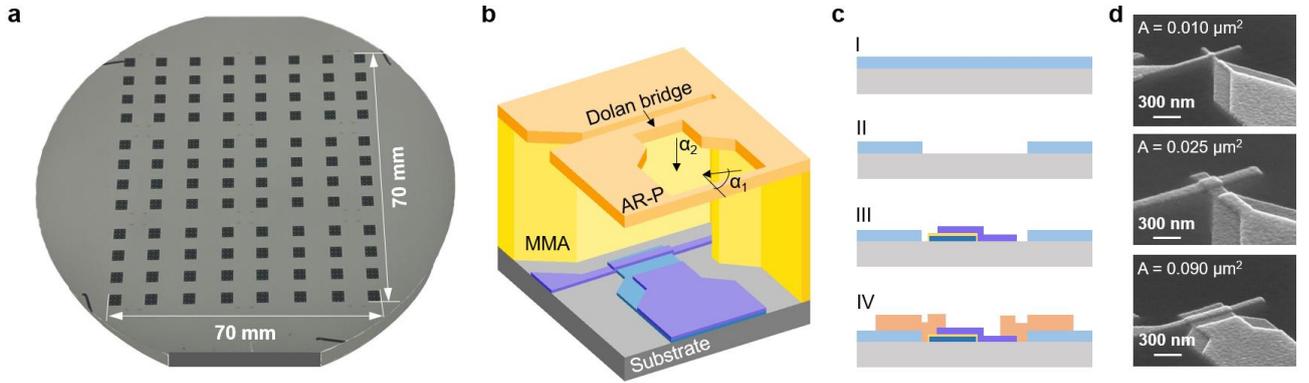

**Figure 1. (a)** General view of a 4-inch wafer with 96 JJ test matrix inside 70×70 mm$^2$ (49 cm$^2$) working area. Each test matrix combines 10 Josephson junction, bandage structure and Al contact pad. **(b)** A shadow evaporation process scheme. We used standard Dolan bridge technique with evaporation angles $a_1 = 40°$ and $a_2 = 0°$. **(c)** JJ evaporation technological steps: I) Al base layer evaporation, II) ground layer dry etching, III) JJ e-beam litho, evaporation and lift-off, IV) Al-bandage evaporation. **(d)** Scanning electron microscopy of Josephson junctions with 0.010 μm$^2$, 0.025 μm$^2$ and 0.090 μm$^2$ areas.

**Table 1.** State-of-the-art Josephson junction reproducibility experimental results

| Paper | JJ type | Working area | JJ area, μm$^2$ | $CV_{R_N}$, % | $CV_A$, % | Optimization technique |
|---|---|---|---|---|---|---|
| [35], 2023 | Manhattan | 27 cm$^2$ | 0.06 | 4.5 | – | Two-step shadow evap. |
| [31], 2020 | Manhattan | 49 cm$^2$ | – | 3.5 | – | EBL optimization |
| [30], 2024 | Dolan | 18 cm$^2$ | 0.02 – 0.08 | 3.7 – 0.8 | – | – |
| [36], 2023 | Manhattan | 10 cm$^2$ | 0.16 | 2.0 | – | Oxidation optimization |
| [32], 2024 | Manhattan | 52 cm$^2$ | 0.04 | 7.0 | 6.0 | Oxidation optimization |
| This work | Dolan | 49 cm$^2$ | 0.025 – 0.09 | 6.0 – 4.0 | 1.0 – 1.1 | SEBi correction + oxidation |
| | | 25 cm$^2$ | 0.025 – 0.09 | 5.2 – 3.4 | 1.0 – 1.1 | |
| | | 16 cm$^2$ | 0.025 – 0.09 | 4.1 – 2.3 | 1.0 – 1.1 | |

In this paper, we demonstrate a robust Al/AlO$_x$/Al Dolan-bridge Josephson junction fabrication process on 4-inch wafers with preliminary shadow evaporation bias resist mask correction and comprehensive oxidation optimization. We propose the model for two-layer resist mask biasing at a wafer-scale, which allows compensating shadow evaporation JJ dimensions shrinks. It takes into account the changes in evaporation angle over a wafer, resist mask geometry, shadow evaporation process parameters including evaporation source geometry. We use a non-point evaporation source for the model that allows increasing the precision of bias calculation for the wafer of different diameters. According to the model each JJ electrode dimension is corrected for the calculated bias depending on its position (coordinates) on the wafer. We named this wafer topology preparation process Shadow Evaporation Bias Correction (SEBi correction). It results in

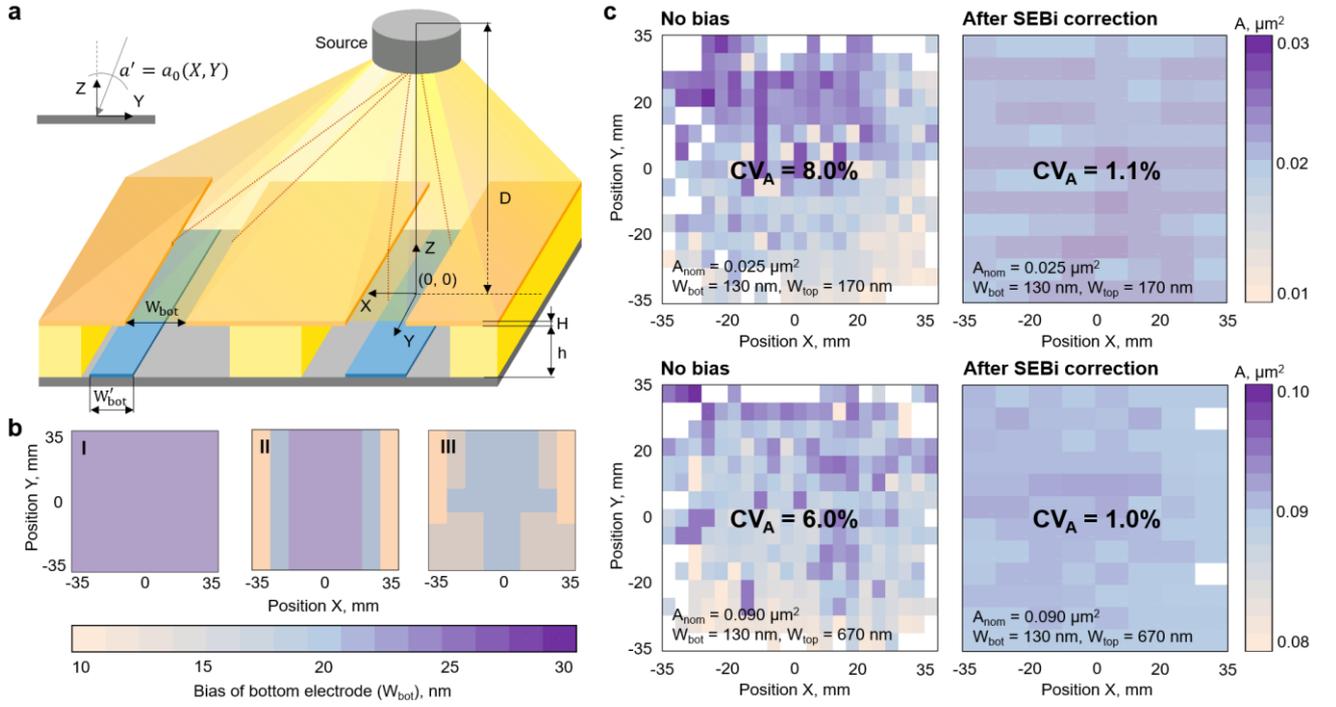

**Figure 2.** (a) The shadow evaporation model geometric representation. Josephson junction dimensions change depending on its wafer position. The proposed model takes into account the source distance $D$, thickness of the top (H) and bottom (h) layers of organic mask, and shadow evaporation regime parameters $\alpha'$. A non-point source model is used to increase the bias calculation precision. (b) Bias corrections of bottom Al electrode dimensions over the wafers. I Bias corrections zones excluding model parameters, II SEBi correction with a point evaporation source, III SEBi correction with a non-point evaporation source. (c) Experimentally measured distribution of Josephson junction areas over the wafers for 130×170 nm$^2$ (0.025 μm$^2$, top map) and 130×670 nm$^2$ (0.090 μm$^2$, bottom map) before (left) and after (right) SEBI correction with a non-point evaporation source.

Josephson junction area variation coefficient ($CV_A$) improvement down to 1.1% for the critical dimensions from 130×170 nm$^2$ to 150×670 nm$^2$ over 70×70 mm$^2$ wafer working area. Next, we investigate JJ oxidation process (oxidation method, pressure and time) and its impact on room temperature resistance reproducibility. We combine both SEBi correction and oxidation best practices for 4-inch wafers improving $CV_{R_N}$ down to 6.0% for 0.025 μm$^2$ JJ area and 4.0% for 0.090 μm$^2$ JJ area for 70×70 mm$^2$ (49 cm$^2$) wafer working area (4.1% and 2.3% for 50×50 mm$^2$ (25 cm$^2$) working area correspondingly). In general, these results can be further improved with junction's post fabrication treatment techniques [33, 34].

## RESULTS

All samples are fabricated using our standard Dolan bridge technology [See Methods]. Figure 1a shows the general view of the test topology. To characterize $R_N$ variation over a wafer, we patterned 96 matrices with the junctions of three different areas as $A = 0.010$ μm$^2$, $A = 0.025$ μm$^2$, and $A = 0.090$ μm$^2$. The surface quality and the junction dimensions were measured by scanning electron microscopy. The $R_N$ dispersion values are calculated using an automatic measurement with a probe station. We optimized both voltage frequency and amplitude to increase measurements reproducibility leading to decrease in $R_N$ standard deviation from 3.2% to 0.5%. A room temperature resistance variation coefficient ($CV_{R_N}$) and Josephson junction area variation coefficient ($CV_A$) are described by:

$$CV_{R_N} = \sigma_{R_N}/R_N \text{ and } CV_A = \sigma_A/A, \tag{1}$$

where $\sigma_{R_N}$ and $\sigma_A$ – standard deviation of the room temperature resistance $R_N$ and Josephson junction area $A$.

A transmon qubit resonance frequency vs room temperature resistance of Josephson junction is described by:

$$hf = \sqrt{\frac{2\Delta\Phi_0}{eR_N}E_C} - E_C, \tag{2}$$

**Shadow evaporation model**

The limitations in Josephson junction parameters reproducibility are originated from both e-beam lithography and shadow evaporation processes [35, 37]. The reason is a diverging metal flow forming during electron-beam evaporation. The conical shape of the metal flow leads to different evaporation angle along the wafer results in changing JJ critical dimensions [21]. Here, we proposed the shadow evaporation process model which eliminates JJ geometry position dependence. The model parameters are the thickness of the top ($H$) and bottom ($h$) layers of two-layer resist mask, substrate-holder tilt angle ($a_0$), JJ resist mask width ($W_{bot}$ and $W_{top}$), structure position on the wafer ($X, Y$) and the distance between crucible and substrate-holder ($D$) [Fig. 2a]. The actual evaporation angle taking into account the structure position on the wafer is described by:

$$\alpha' = a_0(X, Y) = 90 - \cos^{-1}\left(\frac{Y + D\sin(a_0)}{\sqrt{Y^2 + D^2 \pm 2YD\sin(a_0)}}\right). \tag{3}$$

The area of Josephson junction ($A'_{overlap}(X, Y)$) at the position on the wafer with coordinates (X, Y) is described by:

$$A'_{overlap}(X, Y) = W'_{top}(X, Y)W'_{bot}(X, Y). \tag{4}$$

where $W'_{top}(X, Y)$ and $W'_{bot}(X, Y)$ are the width of the top and bottom Al electrodes evaporated at $\alpha'$ actual angle.

Figure 2b (II) shows the required bias corrections of resist mask dimensions over 70×70 mm2 depending on the JJ position on the wafer in case of point-shape evaporation source. It includes both effects of shading of the evaporated Al electrode by the resist mask and changing in the evaporation angle on structure position on the wafer (see Appendix A). One can increase the accuracy of the model, if a more realistic round-shape evaporation source is used [Fig. 2b (III)]. In this case the Al electrode width depending on its position are described as:

$$W'_{bot}(X, Y) = \begin{cases} W_{bot} + \dfrac{c + W_{bot}}{D\cos(a') - h}h, & \text{for } X = 0 \\ W_{bot} + \dfrac{X + c + 0.5W_{bot}}{D\cos(a') - H}(H + h), & \text{for } X \neq 0 \end{cases} \tag{5}$$

Next issue is a bottom Al electrode evaporation process, when a thin Al film grows on a surface and sidewalls of resist mask. It leads to a decreased width of the resist mask at the next step of a top Al electrode evaporation. The grown Al film thickness depending on its position on a wafer is given by the relation:

$$T'(Y) = T_0 (\cos(\alpha'))^2, \tag{6}$$

where $T_0$ – a calibrated film thickness at $a_0 = 0$.

The resulting equation for top electrode dimensions is given by the relation:

$$W'_{top}(X, Y) = \begin{cases} W_{top} - T'(Y) - \dfrac{2D\sin(a') + 2c + W_{top}}{D - h}h, & \text{for } Y = 0 \\ W_{top} - T'(Y) - H\dfrac{D\sin(a') - c - 0.5W_{top}}{D\cos(a') - T' - H - h}, & \text{for } Y \neq 0 \end{cases} \tag{7}$$

We used the relations (5) and (7) to calculate final evaporated width of bottom and top Al electrodes over the wafer. Then we applied SEBi corrections of Josephson junction dimensions over a wafer using the proposed model. Figure 2c shows the experimental results of the SEBi corrections model verification. We demonstrate decreasing in JJ area variation coefficient ($CV_A$) from 7.5% down to 1.1% for 0.025 μm² JJ area and from 6.0% down to 1.0% for 0.090 μm² JJ area.

**Comprehensive oxidation optimization**

Next, we measure JJ room temperature resistance reproducibility over 4-inch wafer in order to trade off our model. The resistance variation coefficient $CV_{R_N}$ for JJ with static oxidation without electrodes dimensions corrections are 16% (for 130×170 nm² junctions) and 14% (for 130×670 nm² junctions) for 70×70 mm² working area [Fig. 3a]. There are two key reasons of wafer-scale JJ resistance nonuniformity, which are JJ area variation (evaporation angle increase from center to wafer edge [Fig. 2a] → additional shading from the top resist → JJ dimensions decrease) and tunnel barrier thickness variation (non-stable poly-Al grain size inside JJ and oxygen concentration over a wafer). Using SEBi correction we almost eliminate the influence of JJ area variation, decreasing $CV_A$ down to 1.1%. It results in decreasing $CV_{R_N}$ from 16% down to 8% (for 130×170 nm²) and from 14% down to 6% (for 130×670 nm² junctions) using the same static oxidation process [Fig. 3b]. One can notice a pronounced gradient in JJ resistance from the left (bigger resistance) to the right (smaller resistance) edge of the wafer.

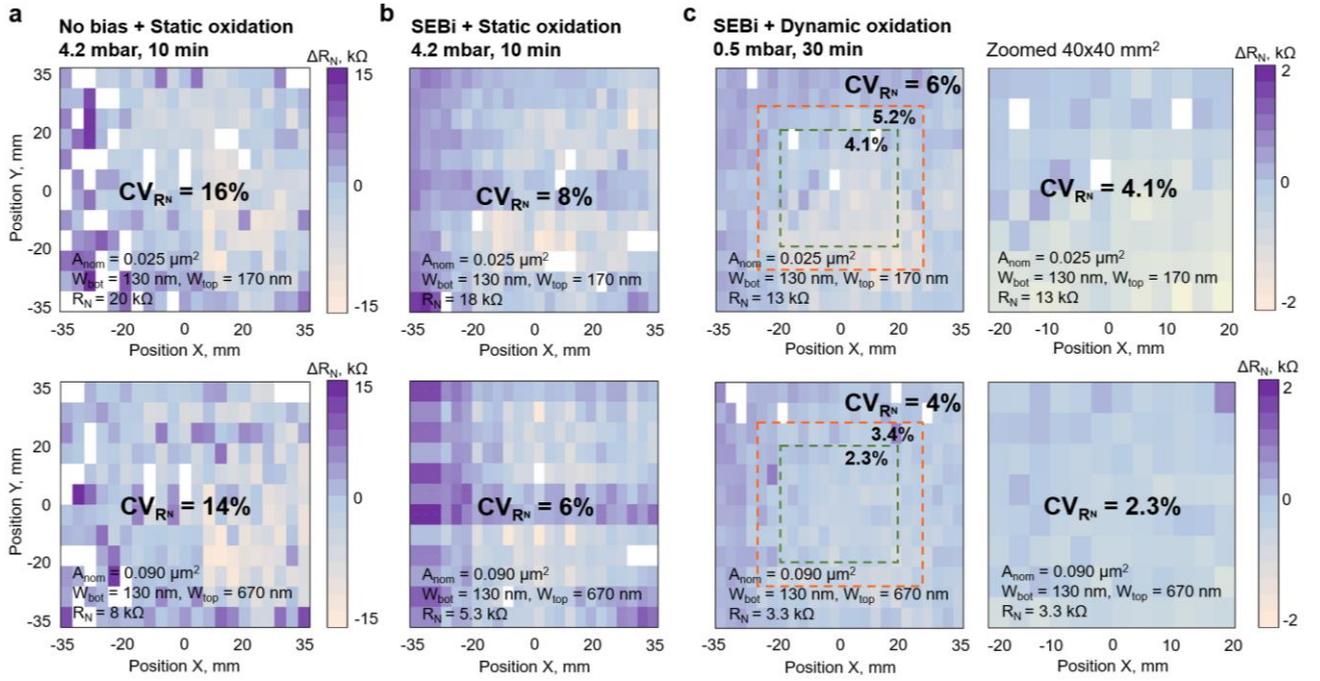

**Figure 3.** The wafer-scale JJ room temperature resistance-map depending on the optimization step for 0.025 μm² and 0.090 μm² junction areas. **(a)** The resistance variation coefficient ($CV_{R_N}$) for JJ with static oxidation without electrodes dimensions corrections is 16% (130×170 nm²) and 14% (130×670 nm²) for 70×70 mm² (49 cm²) working area. **(b)** Using only SEBi correction reduces $CV_{R_N}$ down 8% (0.025 μm² JJ area) and down to 6% (0.090 μm² JJ area) for the same working area. **(c)** The proposed SEBi correction and high-pressure dynamic oxidation allows decreasing the room temperature resistance variation coefficient down 6% (0.025 μm² JJ area) and down to 4% (0.090 μm² JJ area) for 70×70 mm² working area. For 50×50 mm² working area (25 cm², orange square), it is improved down to 5.2% (0.025 μm² JJ area) and 3.4% (0.090 μm² JJ area). For the smallest 40×40 mm² working area (16 cm², green square), it is improved down to 4.1% (0.025 μm² JJ area) and 2.3% (0.090 μm² JJ area).

**Table 2.** Oxidation process optimization parameters and experimental results

| Wafer | 1 | 2 | 3 | 4 |
|---|---|---|---|---|
| Bias | Const | Var | Var | Var |
| Oxidation type | Static | Static | Dynamic | Dynamic |
| Oxidation time | 10 min | 10 min | 30 min | 30 min |
| Oxidation pressure | 4.20 mbar | 4.20 mbar | 0.50 mbar | 0.01 mbar |
| Junction area 0.025 μm² | | | | |
| $CV_A$, % | 8.0 | 1.0 | 0.8 | 1.8 |
| $R_N$, kΩ (meas) | 20 | 18 | 13 | 3.2 |
| $CV_{RN}$, % | 16 | 7.65 | 6.01 | 7.22 |
| $J_c$, μA/μm² | 0.50 | 0.46 | 1.12 | 4.71 |
| Junction area 0.090 μm² | | | | |
| $CV_A$, % | 6.0 | 1.8 | 1.1 | 1.0 |
| $R_N$, kΩ (meas) | 8.0 | 5.3 | 3.3 | 0.9 |
| $CV_{RN}$, % | 14 | 5.79 | 4.02 | 4.64 |
| $J_c$, μA/μm² | 0.51 | 0.47 | 1.15 | 4.56 |

The tunnel barrier growth process can be divided into two oxidation stages: first stage, a very fast almost linear vs time oxide growth up to around 0.5 – 1.0 nm thickness and, second stage, much slower (exponential like) thicker oxidation [38]. For static oxidation an oxygen intake time to reach the desired oxidation pressure, take a considerable part of the overall oxidation time (several minutes). During the oxygen intake, the wafer edge located closer to the oxygen inlet is oxidized faster (first stage) due to enhanced partial (local) pressure and constant refresh of pure oxygen. In this case, the main contribution to a tunnel barrier thickness gradient is the oxygen inlet position in the oxidation chamber. To solve this problem, one can dramatically lower the partial oxygen pressure during oxidation to reduce the influence of oxygen intake time and its concentration uniformity over a wafer. We uses dynamic oxidation with order of magnitude lower oxygen pressure, implementing uniform pure oxygen supply during whole oxidation process (Table 2). The oxygen has faster and more evenly, diffusion at the wafer surface in these conditions leading to high quality and homogeneous tunnel barrier growth [34]. Using SEBi correction with dynamic oxidation we demonstrate decreasing $CV_{R_N}$ from 8% down to 6% (for 130×170 nm$^2$) and from 6% down to 4% (for 130×670 nm$^2$) for 70×70 mm$^2$ working area [Fig. 3c]. For widely used 40×40 mm$^2$ working area it is improved to 4.1% and 2.3% correspondingly. The obtained resistance values correspond to the critical currents range from 10 to 100 nA widely used for high-coherent superconducting qubits fabrication. The proposed technique is also applicable for JJ fabrication with high critical currents (more than 0.5 μA), which can be used for Josephson parametric amplifiers and highly efficient superconducting quantum memory [39].

**DISCUSSION**

Motivated by the challenging task of a high-uniformity Josephson junction fabrication for scalable quantum computing, we undertook a systematic study to identify the reasons and sources of room temperature resistance variation. In this paper, we demonstrate a robust Al/AlOx/Al Dolan-bridge Josephson junction fabrication process on 4-inch wafers with preliminary shadow evaporation bias resist mask correction and comprehensive oxidation optimization. We introduce topology correction model for two-layer resist mask biasing at a wafer-scale, which takes into account a round-shape evaporation source geometry. The proposed Shadow Evaporation Bias (SEBi) corrections allows us increasing the precision of bias calculation for the wafer-scale Josephson junction dimensions calculations. Using SEBi correction we almost eliminate the influence of JJ area variation, decreasing $CV_A$ down to 1.1% for the critical dimensions from 130×170 nm$^2$ to 130×670 nm$^2$ over 70×70 mm$^2$ wafer working area. This step made it possible to separate the influence on JJ resistance variation from different fabrication processes (e-beam lithography, shadow evaporation and oxidation). Using only SEBi correction reduces $CV_{R_N}$ from 16% down to 8% (0.025 μm$^2$ JJ area) and from 14% down to 6% (0.090 μm$^2$ JJ area) for 70×70 mm$^2$ working area. We combine both SEBi correction and high-pressure dynamic oxidation for 4-inch wafers improving $CV_{R_N}$ down to 6.0% for 0.025 μm$^2$ JJ area and 4.0% for 0.090 μm$^2$ JJ area for 70×70 mm$^2$ (49 cm$^2$) wafer working area. For 50×50 mm$^2$ working area (25 cm$^2$) it is improved down to 5.2% (0.025 μm$^2$ JJ area) and 3.4% (0.090 μm$^2$ JJ area). For the smallest 40×40 mm$^2$ working area (16 cm$^2$) it is improved down to 4.1% (0.025 μm$^2$ JJ area) and 2.3% (0.090 μm$^2$ JJ area). Based on the simulation results we show, that the highest JJ reproducibility achieved with a small evaporation angle (before 45°) and small thickness of the top layer of the organic mask. We assume, that the next steps in $CV_{R_N}$ improving are JJ metal-substrate interface optimization and post-fabrication methods.

**METHOD**

**Sample fabrication.**

All the samples are fabricated on 4-inch high-resistivity silicon wafers (> 10000 Ω-cm). First, the substrate was cleaned in Piraniha solution at 80C, followed by dipping in 2% hydrofluoric bath. Second, 120 nm Al film was deposited by the electron-beam evaporation method. Contact pads were defined using a laser direct-writing lithography and wet etch. Josephson junctions are evaporated using our optimized Dolan bridge technology [37, 40] with ultra-low Al electrodes edge roughness. The substrate is spin-coated with a resist bilayer (500 nm MMA-EL9 copolymer and 100 nm AR-P 6200 (CSAR) resist) and exposed with 50 keV electron-beam lithography. The development is performed in a bath of Amylacetate followed by rinsing in IPA for the top layer and IPA:DiW solution for the MMA copolymer. Descum process is carried out after development to clean the substrate from organic residues. Then the junction electrodes are shadow-evaporated in an ultra-high vacuum deposition system. First evaporated Al junction electrode is 25 nm thick and second is 45 nm. We investigate how the oxidation conditions of the tunnel barrier – the oxidation type, time and pressure – affect the JJ room temperature resistance $R_N$. The deposition rate was chosen to provide a minimum for both the root mean square surface (RMS) roughness and the line edge roughness (LER) of the bottom electrode [37]. Then aluminum bandages are defined and evaporated using the same process as for the junctions with an in-situ Ar ion-milling to provide good electrical contact of the junction with the base layer. Lift-off is performed in a bath of N-methyl-2-pyrrolidone with sonication at 80°C and rinsed in a bath of IPA with ultrasonication.

**Author contributions**

D.A.M., N.D.K, D.O.M. and I.A.R. conceptualized the ideas of the project. D.A.M., N.D.K., A.A.S. and M.I.T. fabricated experimental samples and discussed results. D.A.M. and A.S.S. performed morphology characterization. N.D.K. performed shadow evaporation simulation. N.S.S., E.I.M and A.R.M. conducted the electrical characterization of the experimental samples. D.A.M., N.D.K, D.O.M. and I.A.R. analyzed the experimental data and discussed the results. D.A.M., N.D.K, D.O.M. and Y.V.P. prepared writing-original draft. I.A.R. reviewed and edited the manuscript. I.A.R. supervised the project. All authors analyzed the data and contributed to writing the manuscript.

**Acknowledgements**

Technology was developed and samples were fabricated at the BMSTU Nanofabrication Facility (Functional Micro/Nanosystems, FMNS REC, ID 74300).

**Competing interests**

The authors declare no competing interests.

**Data availability**

All data generated or analysed during this study are included in this published article [and its supplementary information files]

**Correspondence** and requests for materials should be addressed to I.A.R.